\input{aipcheck}

\documentclass[
    ,final            
  ]
  {aipproc}

\layoutstyle{6x9}

\def\be{\begin{equation}}
\def\ee{\end{equation}}
\def\bea{\begin{eqnarray}}
\def\eea{\end{eqnarray}}

\newcommand{\lsim}{\raise.3ex\hbox{$<$\kern-.75em\lower1ex\hbox{$\sim$}}}

\begin{document}

\title{Confinement Driven by Scalar Field in 4d Non Abelian Gauge Theories}

\classification{11.25.Mj, 12.90.+b, 12.38.Aw, 12.39.Pn}
\keywords{dilaton, confinement, quark potential}

\author{Mohamed Chabab}{
  address={LPHEA, Physics Department, Faculty of Science Semlalia, Cadi-Ayyad
  University, 40000-Marrakech, Morocco,\\ Email:mchabab@ucam.ac.ma}
}

\begin{abstract}
We review some of the most recent work on confinement in 4d gauge theories
with a massive scalar field (dilaton).  Emphasis is put on the derivation of confining
analytical solutions  to the Coulomb problem versus dilaton effective
couplings to gauge terms. It is shown that these effective theories can be
relevant to model quark confinement and may shed some light on confinement
mechanism. Moreover, the study of an interquark potential derived from Dick
Model, in the heavy meson sector, proves that phenomenological investigation
of this mechanism is more than justified and deserves more efforts.
\end{abstract}

\maketitle


\section{Introduction}
Full Understanding of the QCD vacuum structure and color confinement mechanism are
still lacking. Despite enormous amount of work performed over more than thirty
years, particularly in lattice simulations of QCD, direct derivation of confinement from first
principles remain still elusive, and there is no totally convincing proposal
about its generating mechanism. On the other hand, it is known that the vacuum
topological structure of theories with dilaton fields is drastically changed
compared to the non dilatonic ones \cite{CT}. Therefore much about confinement might be
learned from such theories, particularly  string
inspired ones. Indeed the appearance of fundamental scalars with direct
coupling to gauge curvature terms in string theories offers a challenge with
attractive implications in four-dimensional gauge theories.  \footnote{The dilaton is an hypothetical scalar particle predicted by string theory and
  Kaluza-Klein type theories. In string theory, its expectation
  value  probes  the strength of the gauge
coupling \cite{GSW}.} Besides, color confinement
can be signaled through the behavior of the interaction potential at large distances. In this context,
it was suggested in \cite{dick} that an effective coupling of a
massive dilaton to the 4-dimensional gauge fields may provide an
interesting mechanism wich accomodate both the Coulomb and confining
phases. The derivation performed in \cite{dick, Ch1} suggest a new scenario to
generate color confinement which may be considered as a challenge to the
mechanism based on monopole condensation.

The outline of this contribution is as follows. In the next section, We describe
the influence of the dilaton on a low energy gauge theorie and look into the
problem how dilatonic degrees of freedom modifies Coulomb potential and how
transition to a confinining phase occurs. Then, we review several recent work
by presenting the corresponding effective coupling functions used. We brifly
comment on the analytic solutions of the field equations and their confinement features.
Also, it seems to us more than justified to dedicate some efforts to
phenomenological investigations. We summarize the results obtained from study of Dick interaction potential in the heavy
quarkonium systems.

\section{The model}

The imprint of dilaton on a 4d effective nonabelian gauge theory is described
by a Lagrangian density:

\begin{equation}
{\cal L}({\phi},A)=
-\frac{1}{4F({\phi})}{G_{{\mu}{\nu}}^a}{G^{{\mu}{\nu}}_a}
+\frac{1}{2}\partial_\mu \phi \partial^\mu \phi  -V(\phi) +J_a^\mu
A_\mu^a
\end{equation}

 where $V(\phi)$ denotes the non perturbative dilaton potential
and $G^{\mu \nu}$ is the standard field strength tensor of the theory.
$F(\phi)$ is the coupling function depending on the dilaton
field. Several forms of $F(\phi)$ have been proposed in literature. The most
popular one $F(\phi)=e^{-k\frac{\phi}{f}}$ occured in
string theory and Kaluza-Klein theories\cite{GSW}.
 
The problem of the Coulomb gauge theory augmented
with dilatonic degrees of freedom in (1) is analyzed as follows:\\
First, we consider a point like static Coulomb source which is
defined in the rest frame by the current:

\begin{equation}
J_a^\mu =g \delta (r) C_a \nu_0^\mu =\rho_a \eta_0^\mu
\end{equation}
where $C_a$ is the expectation value of $SU(N_c)$ generator.

The field equations emerging from the static configuration (2) are given by:

 \begin{equation}
 \left[ D_\mu , F^{-1} (\phi ) G^{\mu\nu}\right] = J^\nu
\end{equation}

 and

\begin{equation}
 \partial_\mu \partial^\mu \phi = -\frac{\partial
 V(\phi)}{\partial\phi}-\frac{1}{4} \frac{\partial F^{-1}(\phi)}{\partial
 \phi}G_a^{\mu\nu}G_a^{\nu\mu}
 \end{equation}

At this stage, by seting $G_a^{0i} = E^i \chi_a =-\nabla^i \Phi_a$, after
some algebra, we derive the chromo-electric field: 

\begin{equation}
E_a=\frac{Q^a_{eff}(r)}{r^2}
\end{equation}

where the effective charge is defined by
$$Q^a_{eff}(r)=\left(g\frac{C_a}{4\pi}\right) F(\phi(r))$$. 
From Eq(5), we learn that it is the running of the
effective charge that makes the potential stronger than the
Coulomb potential. In other words, Coulomb spectrum is recovered if the effective charge did not run.  

Thereby the interquark potential reads as \cite{Ch1},

\begin{equation}
U(r)=2\widetilde{\alpha}_s \int \frac{F(\phi(r))}{r^2} dr
\end{equation}

with $\alpha_s = \frac{g^2}{4\, \pi}$ and $\widetilde{\alpha} =\frac{\alpha_s}{8\pi} \left( \frac{N_c -1}{2N_c}\right)$

The formula in is remarkable since it provides a direct relation between the
interquark potential and the coupling function $F(\phi(r))$. Moreover, it
shows that exixtence of a confining phase in the theory in (1) is subject to
the following requirement,

\begin{equation}
\lim_{r\to \infty} r F^{-1}(\phi(r)) = finite
\end{equation}

The main objective is to solve the field equations of motion (3) and (4) and
determine analytically $\phi(r)$ and $\Phi_a(r)$. For this, $F(\phi)$ and
$V(\phi)$ have to be fixed. In the sequel the dilaton potential is set to $V(\phi)
=\frac{1}{2} m^2 \phi$. Below, we will briefly describe the main features of three recent models and
present their solutions.   


\subsection{1. Dick Model}

In this effective theory, Dick used the form: $\frac{1}{F(\phi)} =
\frac{\phi^2}{f^2 + \beta\, {\phi}^2}$ where f represents a coupling scale
characterising the strength of the scalar-gluon coupling. $\beta$ is a
parameter.\\
Then he found for the radial dependance of the dilaton field and the
interquark potential (up to a color factor) \cite{dick}: 
$$ \phi(r)={\pm}\frac{1}{r}\sqrt{\frac{k}{m}+({y_0^2}-\frac{k}{m})exp(-2mr)}$$

$$ V(r) = [\frac{\beta\,  g^2}{4\, \pi\, r}
-gf  \sqrt  \frac{N_c}{2\,(N_c - 1)} ln[e^{2mr} - 1 + \frac{m}{k}y_o^2]$$

With the abbreviation: $k^2=\frac{\alpha_s\, f^2}{8\, \pi}\frac{N_c-1}{N_c}
$ \\

Note that the potential $V(r)$ comes with the required behavior: a first term which accomodates the
Coulomb interaction at short distances, and a second term linearly rising in
the asymptotic regime with a string tension \footnote{In the massless case, $ V(\phi)=0$, solutions of the field equations reduced to:
$\phi(r) = {\pm}\big(\frac{g\,f}{2\pi}\big) \, \sqrt{\frac{N_c -
1}{N_c}}\,r^\frac{-1}{2}$, $V(r) = \frac{g^2\, \beta\, (N_c - 1)}{8\,\pi\,r N_c} -
\frac{f\,g}{2}\sqrt{\frac{N_c - 1}{N_c}}\,r $} \\

$\sigma  \sim g\, m\, f$ which depends on the dilatonic degrees of freedom $m$, $f$.\\

\subsection{2. Cornwall-Soni Model} 

In this model,the glueballs are represented by a massive scalar field $\phi$,
and couple in a non mimimal way to gluons, through $ \frac{1}{F(\phi)} =
\frac{\phi}{f}$ \cite{CS}\\

Analytical Solutions were found for $r  \to \infty $\cite{DF},

$$  \phi(r) =  \Big[\frac{\alpha_s\, f \, (N_c - 1)}{16\, \pi\, m^2\, N_c}\Big]^{\frac{1}{3}} \, 
  r^{-\frac{4}{3}} $$
$$V(r) = -3 \, g \frac{N_c - 1}{2\, N_c} \, \Big[ \frac{g \, f^2 \, N_c
\, m^2}{\pi \, (N_c - 1)}\Big]^\frac{1}{3}r^{\frac{1}{3}}$$

These formulas show that their model provides confinement of quarks detected
through an interaction potential propotional to $r^{1/3}$ at large
distances and considered by the authors as non perturbative correction to the
Coulomb potential.

\subsection{3. Chabab-Sanhaji Model} 

The main aim in this work was to construct a low energy effective field theory from
which some of the popular phenomenological potentials may emerge. For
this,  we used the following coupling function $F(\phi)=\Big( 1 -\beta
\frac{\phi^2}{f^2}\Big)^{-n}$ \cite{chababsanhaji}.\\ 
By substituting $F(\phi)$ in the field equations, they were found too complicated to
integrate analytically. However, as in Cornwall-Soni Model, since the focus is on
the long range behavior of the dilaton field and on how it modifies the
Coulom phase, the analysis is restricted to the infrared region. Thus, the
asymptotic solutions are found to be,   

$$\phi=\Big[ \frac{f^2}{\beta}-\Big(\frac{\beta}{f^2}\Big)^\frac{-n}{n+1}\Big( \frac{2n\alpha_s}{m^2}\Big)^\frac{1}{n+1}\Big(\frac{1}{r}\Big)^\frac{4}{n+1}\Big] $$
and the chromo-electric potential:
$$\Phi_a(r)=-\frac{gC_a}{4\pi}\Big(\frac{2n\alpha_s}{m^2f^2} \Big)^\frac{-4n}{n+1}\frac{n+1}{3n-1}r^{\Big(\frac{3n-1}{n+1}\Big)} $$

We see that the occurrence of confinement depends on the parametre $n$ and our
effective theory can serve to model quark quark confinement when $n\in
\Big[\frac{1}{3}, 1\Big]$. 

On the other hand, we attained the above mentioned objective: by selecting
specific values of $n$, we reproduced the following known interquark potentials
\begin{itemize}
\item   $n=1$ $\Rightarrow$  linear term of Cornwall potential.
\item   $n=11/29$ $\Rightarrow$ Martin's potential \cite{Martin}.
\item   $n=3/5$  $\Rightarrow$ Song-Lin, or Motyka-Zalewski' potential \cite{motyka}.
\item   $n=5/9$ $\Rightarrow$ Turin potential \cite{Turin}.
\end{itemize}

These quark potentials, which gained
credibility only through their confrontation to the hadron
spectrum, are now supplied with a theoretical framework since they can
be derived from a low energy effective theory.


\section{ Phenomenological analysis: Results and Discussion}

The interquark potential resulting from Dick model is quit attractive and
deserves phenomenological investigations. A first study has been performed in \cite{BarakatChabab}.
in the heavy mesons sector. Therein,  the semi-relativistic wave equation  has been solved using Dick
potential. This problem was addressed as in \cite{barakat} where the
shifted-$l$ expansion technique is used (SLET), $l$ is the angular momentum. This method  provides a powerful analytic technique for
determining the bound states of the semi-relativistic wave equation
consisting of two quarks of masses $m_{1}$, $m_{2}$ and total binding meson energy $M$ in
any spherically symmetric potential. It is rapidly converging and handles highly excited states which pose
problems for variational methods \cite{sung}. Moreover, relativistic corrections are
included in a consistent way. \\

 Dick interquark potential reads,
\begin{equation}
V_D(r)=-{4 \over 3}{\alpha_s \over r}+{4 \over 3}gf\sqrt{N_c\over{2(N_c-1)}}\ln[exp(2mr)-1]
\end{equation}

The SLET technique used to obtain results from the theory requires us to
 specify several inputs, namely, $m_{c}$, $m_{b}$, $m$, $f$ and
$\alpha_s$. In our numerical analysis, we set the charm and bottom quark
 masses to the values $m_{c}=1.89$~GeV and $m_{b}=5.19$~GeV. For the QCD
 coupling constant, in contrast to the Lattice potentials which use the same
 effective coupling in the description of heavy quarkonium, we take into account the running of $\alpha_s$, 

\begin{equation}
\alpha_s(\lambda)=\frac{\alpha_s(m_z)}
{1-(11-\frac{2}{3}n_f)[\alpha_s(m_z)/2\pi]ln(m_z/\lambda)},
\end{equation}

 where the renormalization scale is fixed to  $\lambda=2 \mu$, with $\mu$ is the reduced mass,
\begin{equation}
\mu=\frac{m_1m_2}{m_1+m_2},
\end{equation}
 
Thus, combination of the leading order formula (9) and the world experimental value $\alpha_s(m_z)=0.118$ yields,
\begin{equation}
\alpha_s (charmonium)=0.31,  \qquad
\alpha_s (bottomonium)=0.20 ,  \qquad
\end{equation}
 while $\alpha_s=0.22$ for the $b\bar{c}$ quarkonia. 
On the other hand, the interquark potential parameters  $m$ and $f$ are
treated as being free
in our analysis and are obtained by fitting the spin-averaged $c\bar{c}$ and $b\bar{b}$ boundstates. 
An excellent fit with the available experimental
data can be seen to emerge when the following values are assigned,
\begin{equation}
m=57~ MeV \qquad gf\sqrt{\frac{N_c}{2(N_c-1)}}=430~ MeV. 
\footnote{if we adopt the usual number $0.18 GeV^2$ for the string tension, the dilaton mass will be shifted to a value about $158 MeV$.} \qquad
\end{equation}

\begin{table}[ht]
\caption{Calculated mass spectra (in units of GeV) $M_{n\ell}$ of
  $c\bar{c}$ boundstates from  Dick inerquark potential \protect\cite{BarakatChabab} }
\begin{tabular}{|c|c|c||c|c|c|} \hline
 State,$n\ell$ &$M_{n\ell}$, SLET &$M_{n\ell}$, Exp.&
State,$n\ell$ &$M_{n\ell}$, SLET &$M_{n\ell}$, Exp. \\ \hline
1S &3.073  &3.068  &1P&3.546 &3.525\\
2S & 3.662 &3.663  &2P&3.871 &-  \\
3S &4.027  &4.028  &1D&3.787&3.788   \\
\hline
\end{tabular}
\end{table}

\begin{table}[ht]
\caption{Calculated mass spectra (in units of GeV) $M_{n\ell}$ 
of $b\bar{b}$ from Dick interquark potential \protect\cite{BarakatChabab} }
\begin{tabular}{|c|c|c||c|c|c|} \hline
 State,$n\ell$ &$M_{n\ell}$, SLET &$M_{n\ell}$, Exp.&
State,$n\ell$ &$M_{n\ell}$, SLET &$M_{n\ell}$, Exp. \\ \hline
1S &9.450 &9.446  &1P&9.903 &9.900\\
2S & 10.014  &10.013 &2P &10.227 &10.260  \\
3S &10.299  &10.348 &1D &10.129&-   \\
\hline
\end{tabular}
\end{table}

Tables (1,2) list the results of the analysis for the spin-averaged energy levels of
interest. In all cases, where comparison with experiment is possible,
agreement is generally very good. Next step, to check the consistency of our
predictions, we  estimate the bound states energies of the  $b\bar{c}$
quarkonia. These states are  expected to be produced at LHC and
Tevatron. Moreover, they should provide an excellent test to discriminate
between various techniques used to probe nonperturbative properties of
hadrons. In table 3 we show our calculated spectrum. The estimate of the  mass
of the lowest pseudoscalar S-state of the $B_c$ spectra is close to the
experimental value reported by CDF collaboration \cite{cdf}. As to the higher
states masses, they compare favorably with other predictions based on QCD
sum-rules \cite{kis2,chab3} or potential models
\cite{quigg}-\cite{brambilla}. In conclusion, Dick interquark potential (08)
is tested successfully to fit the spin-averaged  quarkonium spectrun. In view of these results, it is quite encouraging to
pursue phenomenological appliation of $V_D(r)$ and other quark potentials
emerging from such low effective gauge theory with dilaton. \\

\begin{table}[ht]
\caption{Calculated mass spectra (in units of GeV) $M_{n\ell}$
of $b\bar{c}$ boundstates from Dick potential \protect\cite{BarakatChabab} }
\begin{tabular}{|c|c|c||c|c|c|} \hline
 State,$n\ell$ &$M_{n\ell}$, SLET &$M_{n\ell}$, Exp.&
State,$n\ell$ &$M_{n\ell}$, SLET &$M_{n\ell}$, Exp. \\ \hline
1S &6.322 &6.40 $\pm0.39$$\pm0.13$ &1P &6.767&-\\
2S & 6.876 &- &2P&7.072&-  \\
3S&7.181 &- &1D&6.994&-   \\
\hline
\end{tabular}
\end{table}

\section{General conclusion}

In summary, We reviewed some of the most recent work on confinement in 4d non
abelian gauge theories with a massive scalar field (dilaton) and  effective
coupling functions to gauge fields.  Analytrical solutions have been found
with confinement feature in the asymptotic regime. Thus, These low energy effective
theories can serve well to model quark confinement. Moreover, by using Dick interquark potential in the heavy quarkonium
sector, we showed that phenomenological investigation of the confinement generating mechanism
suggested by these models is more than justified. Indeed, the obtained results
for charmoniun and bottomonium fit well
experimental data when the dilaton mass is given a value about 57 MeV. Also,
for $B_c$ system, we found that the S-state energy level is close to the value
reported by CDF collaboration, while those of excited states agree favorably
with  predictions of other theoretical works.
On the other hand, This analysis allows a test to the physics beyond the standard model in relation
to hadron spectroscopy. Indeed,  as a by-product, the estimate of the dilaton
mass lies in the range of values proposed in \cite{gasperini,cho}. This
determination may shed some light on the search of the dilaton since, as
suggested in \cite{bando,halyo}, the possibility to identify this hypotetical
particle to a fundamental scalar invisible to present day experiments should not be exluded.

\vspace*{-.1cm}
\begin{theacknowledgments}
The author thanks the CICHEPII organizers for the invitation to this nice
Conference.This Work is partially supported by 
the government research program PROTARS III, contract number D16/04. 

\end{theacknowledgments}

\bibliographystyle{aipproc}

\end{document}